\documentclass[aip,jcp,reprint,floatfix,a4paper]{revtex4-1}
\pdfoutput=1
\usepackage{cmap}
\usepackage[T1]{fontenc}
\usepackage{mathptmx}
\usepackage{times}
\usepackage{fixmath}

\DeclareSymbolFont{UPM}{U}{eur}{m}{n}  % ugly hack to get upright partial; there isn't one in the standard ptmx fonts
\DeclareMathSymbol{\partial}{0}{UPM}{"40}

\usepackage{helvet}

\usepackage{booktabs}

\usepackage[pdftex, bookmarks, pdffitwindow=false, pdfstartview=FitH, pdfdisplaydoctitle, colorlinks, plainpages=false, pdftitle={Effects of surface interactions on heterogeneous ice nucleation for a monatomic water model},pdfauthor={Aleks Reinhardt, Jonathan P. K. Doye}, pdfpagelabels, hypertexnames, citecolor={blue},linkcolor={red}, urlcolor={violet}, pdflang={en}, hyperfootnotes=false, breaklinks]{hyperref}

\usepackage{graphicx}
\usepackage{amsmath}

\usepackage{flafter}

\usepackage{geometry}
\usepackage{setspace}

\usepackage[stretch=15,shrink=15,step=1]{microtype}
\SetProtrusion{encoding={*},family={*},series={*},size={6,7}}
              {1={ ,550},2={ ,300},3={ ,300},4={ ,300},5={ ,300},
               6={ ,300},7={ ,400},8={ ,300},9={ ,300},0={ ,300}}
\usepackage[version=3,arrows=pgf]{mhchem}
\usepackage{siunitx}

\usepackage[nodayofweek]{datetime}
\newdateformat{myDate}{\THEDAY\ \monthname[\THEMONTH] \THEYEAR}

\usepackage{mleftright}

\def\avg#1{\ensuremath{\mleft\langle #1 \mright\rangle}}

\newcommand{\pd}[2]{\ensuremath{\left( \frac{\partial #1}{\partial #2}\right)}}
\newcommand{\der}{\ensuremath{\mathrm{d}}}

\newcommand{\refSub}[2]{\hyperref[#2]{\ref{#2}(#1)}}

\begin{document}

\title{Effects of surface interactions on heterogeneous ice nucleation for a monatomic water model}
\author{Aleks Reinhardt}
\author{Jonathan P.~K.~Doye}
\email[Correspondence author. E-mail: ]{jonathan.doye@chem.ox.ac.uk}
\affiliation{Physical and Theoretical Chemistry Laboratory, Department of Chemistry, University of Oxford, Oxford, OX1 3QZ, United Kingdom}
\date{\myDate\today}

\raggedbottom

\begin{abstract}
Despite its importance in atmospheric science, much remains unknown about the microscopic mechanism of heterogeneous ice nucleation. In this work, we perform hybrid Monte Carlo simulations of the heterogeneous nucleation of ice on a range of generic surfaces, both flat and structured, in order to probe the underlying factors affecting the nucleation process. The structured surfaces we study comprise one basal plane bilayer of ice with varying lattice parameters and interaction strengths. We show that what determines the propensity for nucleation is not just the surface attraction, but also the orientational ordering imposed on liquid water near a surface. In particular, varying the ratio of the surface's attraction and orientational ordering can change the mechanism by which nucleation occurs: ice can nucleate on the structured surface even when the orientational ordering imposed by the surface is weak, as the water molecules that interact strongly with the surface are themselves a good template for further growth. We also show that lattice matching is important for heterogeneous nucleation on the structured surface we study.  We rationalise these brute-force simulation results by explicitly calculating the interfacial free energies of ice and liquid water in contact with the nucleating surface and their variation with surface interaction parameters.
\end{abstract}


% % 64.60.Q-        Nucleation
% % 64.70.D-        Solid-liquid transitions
% % 82.60.Nh        Thermodynamics of nucleation
% % 64.60.qe        General theory and computer simulations of nucleation

\maketitle

\section{Introduction}

The way in which ice forms is important in a variety of fields,\cite{Hagen1981, *Toner1990, *Oxtoby1992, *Karlsson1993, *Baker1997, *Sassen2000, *Debenedetti2003, *Zachariassen2004, *Benz2005, *Hegg2009, *Spichtinger2010, *JohnMorris2011, *Murray2011, *Khvorostyanov2012, *BartelsRausch2012, Murray2012} yet our understanding of the process is still far from satisfactory. Indeed, understanding `how ice forms' has recently been identified as one of the top ten open questions in ice science.\cite{BartelsRausch2013} Very pure water can be cooled considerably below its thermodynamic freezing temperature before it freezes. Understanding the mechanisms of homogeneous ice nucleation is a crucial first step in understanding ice formation generally, and it has been studied extensively over the last few years in microscopic simulations;\cite{Matsumoto2002, *Radhakrishnan2003b, *Radhakrishnan2003, *Quigley2008, *Brukhno2008, *Moore2010, *Moore2011, *Moore2011b, *Li2011, *Li2013, *Geiger2013, Reinhardt2012b, Reinhardt2012, Reinhardt2013c, Sanz2013b} however, in practice, most ice formation on Earth takes place heterogeneously, and it is therefore important to try to understand what role the heterogeneous nucleant plays in the freezing process. In particular, gaining an understanding of how the various `dust' particles present in the air affect the formation of ice in clouds could have fundamental implications in the field of atmospheric science.\cite{Heymsfield2011, Murray2012, Marcolli2014}

Unfortunately, much remains undiscovered about the heterogeneous nucleation pathways relevant to cloud science. For example, feldspar has recently been identified as being particularly important for ice nucleation,\cite{Atkinson2013} but it is unclear what the mechanism of feldspar surface nucleation is.  There are a number of fundamental questions about the microscopic details of such processes, and little is really known about them. For example, where does ice nucleate and how? Are planar surfaces sufficient to catalyse nucleation, or do defects and curvature play a major role? That much is still unknown about the nucleation mechanism is perhaps unsurprising, as it is difficult to exercise precise control in experiment: it is often the case that heterogeneous nucleation proceeds on nucleants which proved impossible to remove, and they are therefore often difficult to characterise fully. For a recent review of experimental approaches to heterogeneous nucleation, see Ref.~\citenum{LadinoMoreno2013}.

Computer simulations can provide a route to understanding the microscopic mechanisms that govern heterogeneous ice nucleation without the difficulties of surface characterisation that can plague systematic experimental investigations. Several computer simulations of heterogeneous ice nucleation have been performed so far, including studying nucleation near a vapour interface,\cite{Vrbka2006, *Vrbka2007, *Pluharova2010, Lue2013} on Lennard-Jones and kaolinite surfaces,\cite{Cox2012, Cox2013} on metal surfaces,\cite{Raghavan1991, Carrasco2011, Zhang2013d} in strong electric fields near surfaces,\cite{Yan2011, *Yan2012, *Yan2013} in nanoscale pores,\cite{Moore2010b, *GonzalezSolveyra2011, *Johnston2012} and on graphitic surfaces,\cite{Lupi2014, Lupi2014b, Singh2014} and a considerable degree of insight has already been gained from such work. For example, it has been shown that surface roughness both at the molecular and nano-scale levels\cite{Lupi2014,Singh2014,Nistor2014} appears to decrease the nucleation rate relative to a smooth surface; curvature likewise seems to lead to a reduction in the nucleation rate.\cite{Lupi2014}

A simple approach to understanding the basic physics of heterogeneous nucleation involves multiplying the classical nucleation theory free energy barrier to nucleation by a geometric factor,\cite{Sear2007}
\begin{equation}
f(\theta) = (2+\cos\theta)(1-\cos\theta)^2 / 4,
\end{equation}
where $\theta$ is the contact angle between the wall and the growing crystalline nucleus, which accounts for the changed geometry of the crystalline nucleus relative to the homogeneous case. The contact angle can be related to the interfacial free energies via Young's equation,\cite{Young1805}
\begin{equation}
 \gamma_\text{wall-crystal} + \gamma_\text{crystal-liquid} \cos \theta = \gamma_\text{wall-liquid},\label{eq:youngs-eqn}
\end{equation}
where $\gamma_{ij}$ is the interfacial free energy between phases $i$ and $j$. This contact angle is determined by the interactions between the three pairs of structures, and in general, if the crystal has a favourable interaction with the wall, the angle $\theta$ will be small (this is known as `wetting'), whilst the converse is true if the cluster has a disfavourable interaction with the wall (this is the `drying' regime);\cite{Bonn2001} however, it should be borne in mind that even if the interfacial free energies of the surface interacting with the liquid and the crystal are identical, \textit{i.e.}~$\cos\theta = 0$, $f(\ang{90})=1/2$, and so the free energy barrier is still half that of the corresponding homogeneous nucleation case. Furthermore, what controls the interaction strengths is not necessarily obvious. For example, it has long been assumed that a good heterogeneous nucleant will have a nearly perfect lattice match with ice,\cite{Vonnegut1947, *Vonnegut1949, Turnbull1952, Taylor1993, *Hale1980, *Ward1982} as is the case with silver iodide, which has been used for many years to nucleate ice and reduce the impact of hail storms.\cite{Wieringa2006} Nevertheless, recent simulation work suggests that a lattice match is not necessarily a sufficient criterion, nor indeed is heterogeneous nucleation necessarily fastest on a substrate that has a perfect crystalline lattice match with the nucleating phase.\cite{Cox2012, Mithen2014, *Mithen2014b} Although many of the limitations of classical nucleation theory are widely appreciated,\cite{Oxtoby1998, *Anwar2011b, *Sear2012} the theory has nonetheless been shown to work well in studies of homogeneous ice nucleation,\cite{Reinhardt2013c, Pereyra2011, Sanz2013b} and a similar approach to heterogeneous nucleation using the above equations may provide an alternative means to studying heterogeneous nucleation computationally; namely, it may be easier to compute interfacial free energies than to simulate heterogeneous nucleation directly, particularly for all-atom models of water, for which the crystallisation dynamics can be very slow.

In this work, we look at the heterogeneous ice nucleation behaviour of model flat (atomless) and structured surfaces using the mW model of water. Unlike the simulations of Lupi and co-workers\cite{Lupi2014, Lupi2014b} or Singh and M\"{u}ller-Plathe,\cite{Singh2014} we do not consider particular experimental surfaces, but instead investigate some generic features of heterogeneous nucleation on model surfaces.

\section{Methods}
In the simulations reported here, we have used the mW monatomic model of water proposed by Molinero and Moore,\cite{Molinero2009} which has been shown to provide an excellent description of the thermodynamics and structure of water,\cite{Moore2010} but is much faster than all-atom models of water to simulate, allowing processes to be studied that may not be accessible to simulations using more realistic models of water.

We run hybrid Monte Carlo simulations,\cite{Duane1987, *Heermann1990, *Mehlig1992,*Brass1993} in which short MD simulations replace single particle Monte Carlo moves. We sample in the  isobaric-isothermal ensemble using Monte Carlo volume sampling;\cite{Frenkel2002,Eppenga1984} in simulations with interfaces, we equilibrate the volume in the direction orthogonal to the interface only. To quantify whether water particles are ice-like or not, we use a local order parameter.\cite{Reinhardt2012b}

\subsection{Flat and structured walls}
To account for `generic' interactions of water with surfaces, we first introduce a Lennard-Jones flat wall. The 12-6 Lennard-Jones potential can be integrated in cylindrical polar co-ordinates to give the interaction potential\cite{Wu2010, *Sun2013, Lee1984}
\begin{equation}
 U_\text{fw}(r) = \varepsilon_\text{fw} \left( \frac{2}{15} (\sigma_\text{fw}/r)^9- (\sigma_\text{fw}/r)^3 \right),
\end{equation}
where $r$ is the perpendicular distance from the surface to the particle with which the surface is interacting.

For structured surfaces, we equilibrated a block of mW ice I$_\text{h}$ at the simulation temperature to find the equilibrium lattice parameter. We then took one layer of the basal plane of perfect ice I$_\text{h}$ with the equilibrium lattice parameter and placed it at the top and, in some simulations, the bottom of a simulation cell filled with either ice or liquid water. The interaction of surface particles with bulk water particles is analogous to the mW potential,
where the two- and three-body terms for pairs of particles within the cutoff distance are given by
\begin{equation}
U_2(r_{ij}) = \alpha^{n_{ij}} A\varepsilon\left( B\left[ \sigma/r_{ij}  \right]^4 - 1  \right) \exp\left(  \frac{\sigma}{r_{ij}-a\sigma}  \right)
\end{equation}
and
\begin{equation}\begin{split}
U_3(r_{ij},\,r_{ik},\,\theta_{jik}) &=  \beta^{n_{ijk}} \lambda\varepsilon\left( \cos \theta_{jik} + 1/3  \right)^2 \\ &\qquad\times \exp\left(  \frac{\gamma \sigma}{r_{ij}-a\sigma} +  \frac{\gamma \sigma}{r_{ik}-a\sigma}  \right).
\end{split}\end{equation}
All the parameters\footnote{For reference, $A=\num{7.049556277}$, $B=\num{0.6022245584}$, $a=1.8$, $\gamma=1.2$, $\lambda=23.15$, $\varepsilon/k_\text{B}=\SI{3114.42238}{\kelvin}$ and $\sigma=\SI{2.3925}{\angstrom}$.} are identical to those of mW water,\cite{Molinero2009, Molinero2006} except that, in order to investigate the role of orientational ordering relative to that of a simple lattice matching, the values of $\alpha$ and $\beta$, which are unity in the mW parameterisation, can be varied to give a greater or lesser weight to two- or three-body terms, respectively; $n$ is the number of particles amongst $i$, $j$ and $k$ (as appropriate) that are surface particles. A two- or three-body interaction involving at least one surface particle is considered to be a surface interaction for the purposes of thermodynamic integration, and if all the particles involved are surface particles, then their interaction is not considered at all. In simulations with a rigid structured surface, the $z$-direction of the simulation box is no longer periodic, and just below the structured surface there is therefore a hard wall.

\subsection{Interfacial free energies}\label{subsect-interfacial-free-energies-method}
For an inhomogeneous system with walls, the interfacial free energy is given by the excess Gibbs energy per unit area,\cite{Benjamin2012}
\begin{equation}
\gamma = \frac{G_\text{system with wall} - G_\text{bulk}}{a},
\end{equation}
where `system with wall' refers to the system of interest and `bulk' is the equivalent system with the wall removed. These free energies can be obtained using the thermodynamic integration approach of Benjamin and Horbach.\cite{Benjamin2012,Benjamin2013,*Benjamin2013b} Because we are considering systems in which there is a crystal in contact with a wall, we remark that this interfacial free energy is not equal to the surface stress:\cite{Shuttleworth1950, Frenkel2013} the surface stress also depends on the rate of change of the interfacial free energy with the surface area,\cite{Shuttleworth1950} and, for a solid, the surface structure is changed if it is stretched.\cite{Frenkel2013} To simplify matters, we consider only the interfacial free energy of water in contact with rigid and with structureless walls. The following steps are taken to determine the relevant interfacial free energies:\cite{Benjamin2012}
\begin{enumerate}
 \item We compute the Gibbs energy change $\upDelta G_1$ on the transformation of the bulk system (either liquid water or ice) to a system where periodicity has been switched off in the $z$-direction. This can be obtained by hamiltonian thermodynamic integration\cite{Vega2008} using the potential
 \begin{equation}
  U_1(\lambda) = (1-\lambda) U_\text{periodic} + \lambda U_\text{non-periodic},
 \end{equation}
 where $\lambda$ varies from $0$ to $1$ and $U_\text{periodic}$ and $U_\text{non-periodic}$ are the relevant potential energies with full periodicity and with periodicity only in the $x$- and $y$-directions, respectively. The Gibbs energy change for this transformation is given by\cite{Vega2008, Benjamin2012}
 \begin{align*}
  \upDelta G_1 &=  \int_0^1 \avg{\pd{U_1(\lambda)}{\lambda}}_\lambda \,\der \lambda, \\
  &= \int_0^1 \avg{U_\text{non-periodic}- U_\text{periodic}}_\lambda \,\der \lambda.
 \end{align*}

 \item We then compute the Gibbs energy change $\upDelta G_2$ when a flat, structureless wall is introduced into this non-periodic system via the potential
 \begin{equation}
  U_2(\lambda) = U_\text{non-periodic} + U_\text{fw}(\lambda),
 \end{equation}
 where $U_\text{fw}$ includes all the interactions of particles with the flat wall, and is given by
\begin{equation}
U_\text{fw} (\lambda) =  \lambda^2 \varepsilon_\text{fw} \left(\frac{2}{15} \left(\frac{\sigma_\text{fw}}{r + z}\right)^9 - \left(\frac{\sigma_\text{fw}}{r + z}\right)^3 \right),
\end{equation}
 where $z=(1 - \lambda) \sigma_\text{mW}$ and $\sigma_\text{mW}=\SI{2.3925}{\angstrom}$.
 Note that, following Benjamin and Horbach,\cite{Benjamin2012} we have squared the $\lambda$ dependence, and this should be taken into account when calculating the derivative of the potential with respect to $\lambda$. There is also a $\lambda$ dependence in the denominators of $U_\text{fw}$; this allows us to shift the minimum in the potential gradually away from the wall boundary. We cut and shift the potential at a cutoff of $3\sigma_\text{fw}$, which affects both the potential energy and the derivatives with respect to $\lambda$ in a straightforward way.

 \item Finally, we compute the Gibbs energy change $\upDelta G_3$ on the transformation of this structureless wall to a rigid structured wall by performing an analogous integration in $\lambda$ of the potential
 \begin{equation}
  U_3(\lambda) = U_\text{non-periodic} + (1-\lambda)^2 U_\text{fw}(1) + \lambda^2 U_\text{sw}(\lambda) ,
 \end{equation}
where $U_\text{sw}$ is the potential giving the interaction between bulk particles and surface particles.
\end{enumerate}
The integrals in the above thermodynamic integrations were calculated by non-linear regression fitting of the data points, followed by analytical integration.

To find the interfacial free energy for the liquid in contact with the flat structureless wall, we calculate
\begin{equation}
 \gamma_\text{l-fw} = \frac{G_\text{fw}}{a} - \frac{G_\text{bulk}}{a} = \frac{\upDelta G_2}{a} + \frac{\upDelta G_1}{a}.
\end{equation}
Similarly, for the interfacial free energy between the liquid and the structured wall, we have
\begin{equation}
 \gamma_\text{l-sw} = \frac{G_\text{sw}}{a} - \frac{G_\text{bulk}}{a} =   \frac{\upDelta G_3}{a} + \gamma_\text{l-fw} .
\end{equation}
Equivalent expressions hold for interfacial free energies involving ice.

\section{Results}
\subsection{Flat surface}

\begin{figure}[t]
\centering
\includegraphics{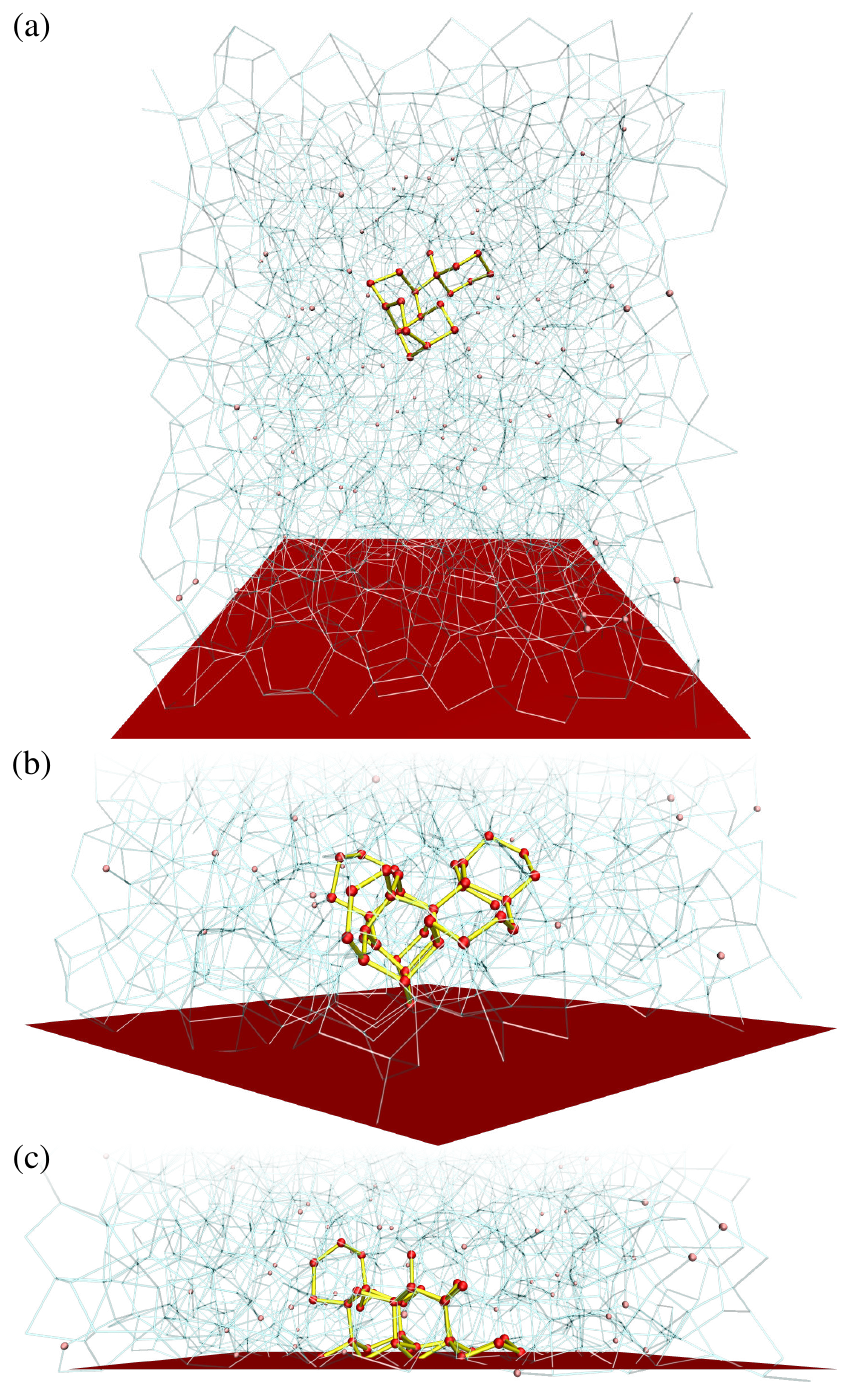}
\caption{(a) A typical cluster forming far away from the flat surface, which is shown in red at $z=\sigma_\text{fw}$. $N=1859$, $T=\SI{210}{\kelvin}$, $p=\SI{1}{\bar}$, $\varepsilon_\text{fw}/k_\text{B} = \SI{186}{\kelvin}$, $\sigma_\text{fw}=\SI{2}{\angstrom}$.  (b) A cluster growing away from the surface in an umbrella sampling simulation. $N=1773$, $T=\SI{220}{\kelvin}$, $p=\SI{1}{\bar}$, $\varepsilon_\text{fw}/k_\text{B} = \SI{177}{\kelvin}$, $\sigma_\text{fw}=\SI{2}{\angstrom}$. (c) A shrinking pre-formed cluster. $N=2461$, $T=\SI{220}{\kelvin}$, $p=\SI{1}{\bar}$, $\varepsilon_\text{fw}/k_\text{B} = \SI{246}{\kelvin}$, $\sigma_\text{fw}=\SI{2}{\angstrom}$. In (b) and (c), only the region of the simulation box close to the surface is shown. Water molecules classified as ice which belong to the largest crystalline cluster are shown in red and connected with yellow bonds; other ice molecules are shown in pink. Unless otherwise connected, all water molecules are connected with thin cyan bonds if they lie within \SI{3.5}{\angstrom} of each other.}\label{fig-mw-flatsurface-nucl-snapshots}
\end{figure}

We ran simulations with a flat surface placed at one end of the simulation box for a range of $\varepsilon_\text{fw}$ and $\sigma_\text{fw}$. The other end of the simulation box comprised a surface with the same potential, but where $\varepsilon_\text{fw}$ was so small that, for all intents and purposes, it was a hard surface. At sufficiently low temperatures, the systems crystallise, but predominantly homogeneously. A typical small cluster is shown in Fig.~\refSub{a}{fig-mw-flatsurface-nucl-snapshots}. There were very few nucleation events near the surface, and when small clusters did form, they quickly fell apart. For some of these surface clusters, we started additional umbrella sampling\cite{Torrie1977} simulations to try to force them to grow; however, we sometimes observed that small (unbiassed) clusters could form in the bulk even when we were biassing the surface cluster to grow. When the surface cluster did grow, it did not do so at the surface, but rather far from the surface: that is, the contact angle was very large (Fig.~\refSub{b}{fig-mw-flatsurface-nucl-snapshots}). We also attempted to see if `epitaxial' growth would be favoured by placing pre-formed clusters with various faces exposed to the surface (for example, Fig.~\refSub{c}{fig-mw-flatsurface-nucl-snapshots}). However, the clusters that we placed on the surface melted relatively quickly, demonstrating that the critical nucleus on the surface must be very large. Finally, for very high values of $\varepsilon_\text{fw}$, a high density layer of water molecules formed at the minimum in the surface potential, and this does not favour heterogeneous nucleation, as ice has a lower density than the liquid.

While the fact that more ice nuclei are formed spontaneously in the bulk than near the surface does not necessarily mean that homogeneous nucleation is favoured over heterogeneous nucleation with this type of surface, as the free energy for the formation of a \textit{critical} nucleus size could still be lower for the surface clusters, the difficulty in driving nucleation near a surface relative to driving it in the bulk with umbrella sampling does suggest that heterogeneous nucleation is disfavoured in this system. In general, it is surprising when a heterogeneous nucleation process is slower than a homogeneous one. One would intuitively expect that nearly any type of surface would significantly enhance the nucleation rate, because one would anticipate that the crystal, with its well-defined crystal planes, would be more compatible with a planar surface, and further that the crystal will have a stronger interaction with the surface because of its (usually) higher density. It is worth reiterating that, in the framework of heterogeneous nucleation theory, the surface need not have a favourable interaction with the crystal, but simply a more favourable interaction than with the liquid.

\begin{table}[t]
\caption{Free energies at \SI{273}{\kelvin} and \SI{1}{\bar} following each step of the thermodynamic integration and the resulting interfacial free energies. The flat wall (fw) parameters are $\varepsilon_\text{fw}/k_\text{B} = \SI{300}{\kelvin}$, $\sigma_\text{fw} = \SI{4.2}{\angstrom}$. The structured wall (sw) parameters are $\alpha=1$, $\beta=1$, 360 surface particles. All values are reported in units of \si{\milli\joule\per\metre\squared}.}\label{table-gammas}
\centering
\begin{tabular}{l S[table-format=2.1, separate-uncertainty=true, table-figures-uncertainty=1]  S[table-format=-2.1] S[table-format=-3.1] S[table-format=2.1] S[table-format=-3.1] }\toprule
 $x$  & {$\upDelta G_1/a$}      &  {$\upDelta G_2/a$} & {$\upDelta G_3/a$} & {$\gamma_\text{$x$-fw}$} & {$\gamma_\text{$x$-sw}$} \\  \midrule
ice & 99.0\pm 0.3  & -39.5 & -177.2  &  59.5 & -117.7 \\
liquid & 67.6 & -28.1 & -113.5 & 39.4 & -74.0  \\ \bottomrule
\end{tabular}
\end{table}

From the simulations performed here, we find that the flat wall does not facilitate nucleation, and in retrospect, perhaps we ought not to have been surprised by this at all. Firstly, the density of liquid water is greater than that of ice, and an attractive surface will thus favour the liquid phase. Secondly, the mW potential imposes an energetic penalty for non-tetrahedral triplets, and by removing neighbours at one end (such as at a surface), the penalty for non-tetrahedral bond angles is decreased, and this reduction in tetrahedrality favours the liquid phase over ice. In this respect, one should exercise care when using the mW potential near a surface that was not specifically parameterised to account for mW's three-body potential term.

To quantify the behaviour we have observed in brute-force simulations, we have calculated the interfacial free energies of liquid water and ice in contact with the flat surface by employing the method outlined in subsection~\ref{subsect-interfacial-free-energies-method}. The interfacial free energies at \SI{273}{\kelvin} are summarised in Table~\ref{table-gammas}. These allow us to calculate the contact angle $\theta$ using Young's equation (Eqn~\eqref{eq:youngs-eqn}). If we assume that the classical nucleation theory result $\gamma_\text{crystal-liquid} = \SI{26.2}{\milli\joule\per\metre\squared}$ obtained from a free energy profile for the mW model of water is a reasonable estimate,\cite{Reinhardt2012} then we find that the contact angle is approximately \ang{140} for the flat wall. Such a flat wall is thus not helpful in facilitating ice nucleation: indeed, the liquid phase is preferred, because the wall has a stronger interaction with the higher density liquid phase, which is consistent with what we observe in brute-force simulations.

\subsection{Structured surface}
\subsubsection{Rigid ice surface}

\begin{figure}[t]
\centering
\includegraphics{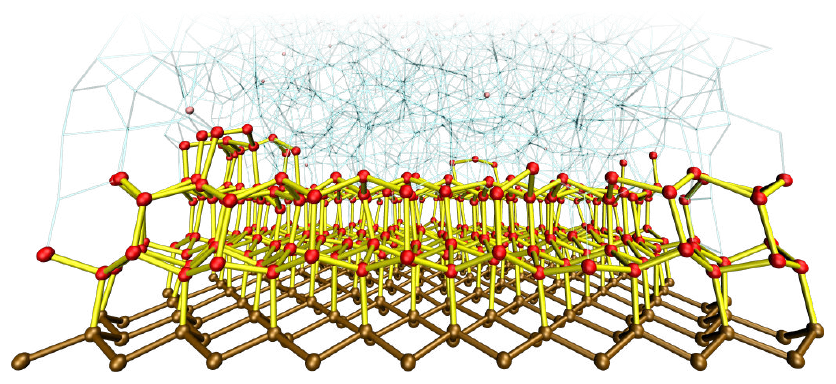}
\caption{Typical ice growth from a structured surface, where in this case the surface particle interactions with the moving particles are identical to the interactions between the moving particles, and growth is in the wetting regime. $N=2160$ (of which 180 are fixed surface particles), $T=\SI{273}{\kelvin}$, $p=\SI{1}{\bar}$, $\alpha=\beta=1$. Only the region of the simulation box close to the surface is shown. The colour scheme is the same as in Fig.~\ref{fig-mw-flatsurface-nucl-snapshots}, with the structure of the ice-like surface shown in brown.}\label{fig-mw-atomicsurf-fullInteraction}
\end{figure}

\begin{figure}[tbp]
\centering
\includegraphics{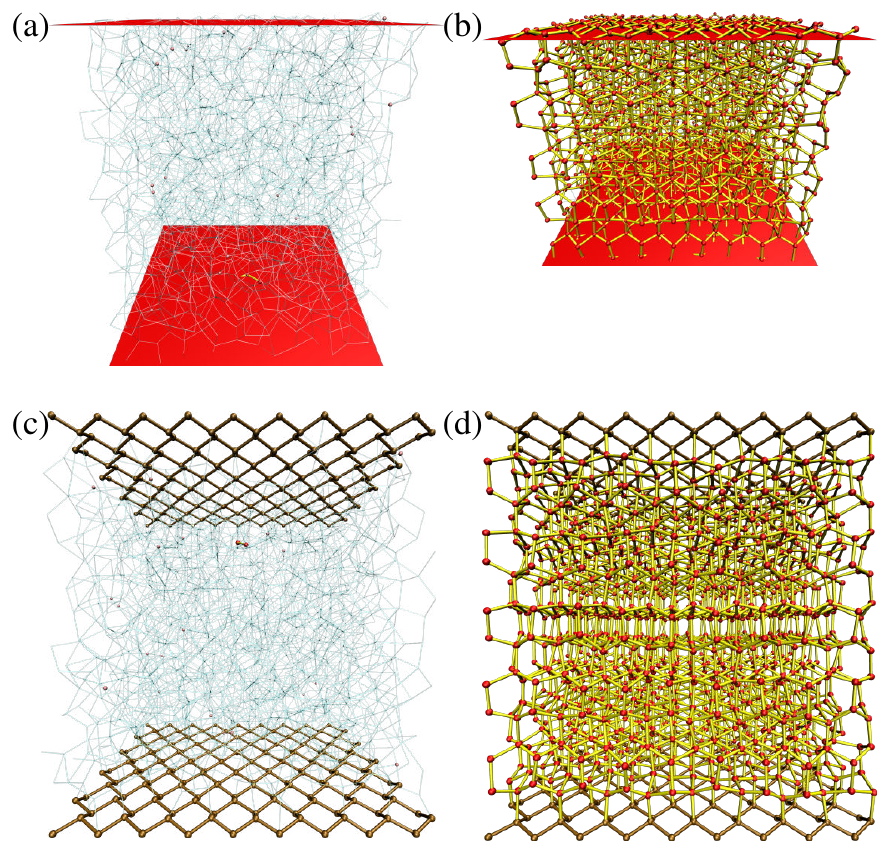}
\caption{Typical configurations for (a), (c) liquid water and (b), (d) ice from step 2 ((a), (b), $\lambda=1.0$) and step 3 ((c), (d), $\lambda=0.61$) of the thermodynamic integration. The interaction parameters for the top and bottom surface are identical in each case. $T=\SI{273}{\kelvin}$, $p=\SI{1}{\bar}$. In (a) and (b), $\varepsilon_\text{fw}/k_\text{B}=\SI{300}{\kelvin}$, $\sigma_\text{fw}=\SI{4.2}{\angstrom}$; in (a), $N=1859$ and in (b), $N=1800$. In (c) and (d), $\alpha=1.0$, $\beta=1.0$, $N=2160$, of which 360 are fixed surface particles.}\label{fig-mw-interfacialEnergy-TI-snapshots}
\end{figure}

Simulations involving a structured wall with $\alpha=\beta=1$ are rather similar to direct coexistence simulations,\cite{Ladd1977, Fernandez2006, *Carignano2007, *Rozmanov2012c} and so when the temperature is lower than the freezing point, rapid freezing is expected in the `wetting' regime. This is indeed what we observe in brute-force simulations; a simulation snapshot is shown in Fig.~\ref{fig-mw-atomicsurf-fullInteraction}

As before, we can also calculate the interfacial free energies of the liquid and the ice interacting with the structured surface; several snapshots from the relevant stages of the interfacial free energy calculation are shown in Fig.~\ref{fig-mw-interfacialEnergy-TI-snapshots}. One potential difficulty in calculating interfacial free energies of the liquid, in particular when in contact with the structured surface, is the fact that the liquid will undergo facile crystallisation; to help avoid this, we calculate the interfacial free energies close to the coexistence temperature. The interfacial free energies at \SI{273}{\kelvin} are summarised in Table~\ref{table-gammas}. As we did with the flat surface above, we can calculate the contact angle $\theta$ using Young's equation. Following the same procedure as above, we find that $\cos \theta > 1$ for the structured surface. The structured wall is thus in the wetting regime and is very good at nucleating ice, as we have seen in the brute-force simulations.

That $\gamma_\text{ice-sw}$ (Table~\ref{table-gammas}) is so negative may seem surprising, because the structured wall is in fact just a block of ice itself. However, the structured wall is made up of `perfect' ice with the correct lattice parameter, whereas the ice in bulk simulations is fairly distorted at these temperatures by thermal motion, as the simulation is just below coexistence. Thus ice appears to favour the rigid structured wall over itself.

To ensure that our interfacial free energy data are reasonable, we have run several `consistency' checks, as suggested by Vega and co-workers,\cite{Vega2008} to verify our free energy calculations. In particular, we have calculated the equivalent results to those reported in Table~\ref{table-gammas} for the liquid at \SI{310}{\kelvin} and for ice at \SI{255}{\kelvin}, and verified that using standard thermodynamic integration along an isobar, we obtain the same results. Whilst we appreciate that this is not a rigorous test of the method, the consistency in the results obtained in these different ways suggests that our implementation of the interfacial free energy calculation is correct. The method itself has been verified extensively by Benjamin and Horbach.\cite{Benjamin2012, Benjamin2013, *Benjamin2013b}

\begin{figure}[tbp]
\centering
\includegraphics{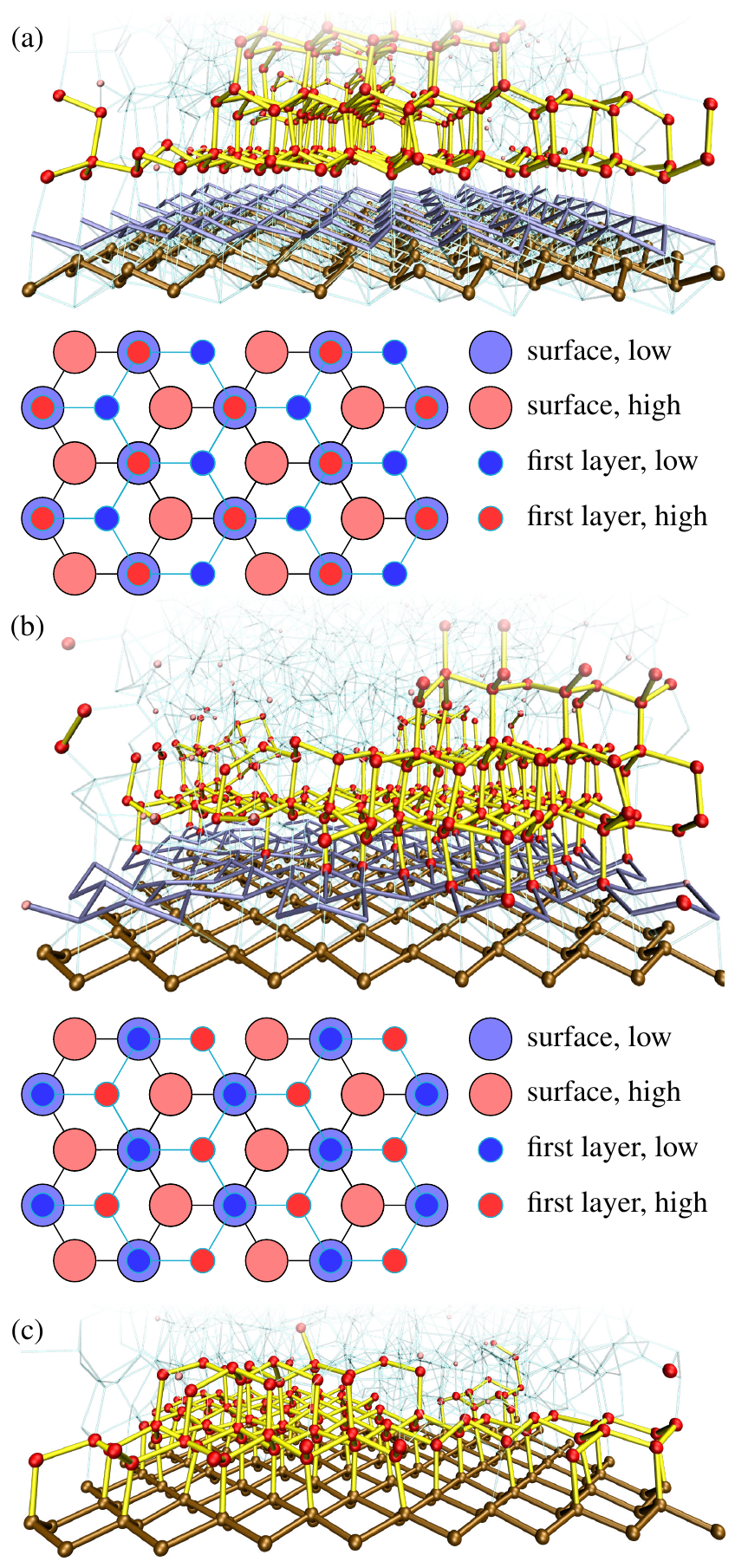}
\caption{Early structured surface nucleation snapshots.  $N=2294$, of which 160 are fixed surface particles; $T=\SI{220}{\kelvin}$, $p=\SI{1}{\bar}$. Only the region of the simulation box close to the surface is shown in each case. In (a), liquid molecules penetrate the surface; ice can then grow on top of the underlying structure. $\alpha=1.0$, $\beta=0.25$. In (b), $\alpha=0.3$, $\beta=0.25$ and in (c), $\alpha=1.5$, $\beta=1.0$. In (a) and (b), the first `movable water' layer is connected with steel blue bonds. In (a) and (b), sketches of the idealised locations of water molecules in the surface and first `movable water' layers are also shown in plan view. In this schematic representation, the blue-coloured molecules have a smaller $z$-co-ordinate value than the red-coloured ones within the same layer.}\label{fig-mw-atomic-nucl-snapshots}
\end{figure}

\subsubsection{Relative effects of two- and three-body interactions}
When considering the nucleation behaviour on a structured surface, a key question to address is what the main driving force for nucleation to occur is: is all that we require simply a lattice match, or is the imposition of orientational order through three-body interactions also important? With a structured surface, this nucleation behaviour depends strongly on the two- and three-body strength parameters $\alpha$ and $\beta$. For example, if $\beta$ is very small, then there is an insufficient penalty for non-tetrahedral arrangements and a densification at the surface results in the direction orthogonal to the plane of the surface, which allows water molecules to form an additional layer in the hollows of the surface structure. This arrangement of molecules can still nucleate ice, but it is actually the `movable' water molecules rather than the rigid ones that serve as the starting point for ice growth (Fig.~\refSub{a}{fig-mw-atomic-nucl-snapshots}).

\begin{figure}[tbp]
\centering
\includegraphics{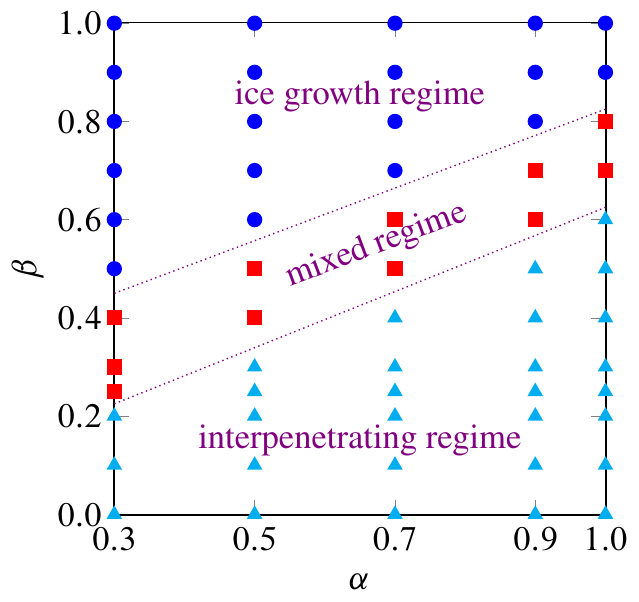}
\caption{A schematic diagram showing the principal growth mode as a function of the two- and three-body strength parameters $\alpha$ and $\beta$. $T=\SI{220}{\kelvin}$. The circles, squares and triangles correspond to particular brute-force simulations, with symbols mapped to the appropriate nucleation regime; the dotted lines, denoting the boundaries of each regime, are guides to the eye only.}\label{fig-mw-atomic-alphabeta-schematic}
\end{figure}

A decrease in $\alpha$ when $\beta$ is very small can reduce the penetration of water molecules into the surface, but an intervening nominally non-ice-like layer (with too many neighbours for each water molecule) forms between the surface and the nucleating ice (Fig.~\refSub{b}{fig-mw-atomic-nucl-snapshots}). By contrast, increasing the two-body strength within reason when $\beta=1$  does not affect the structure that forms significantly, as the three-body terms prevent any particularly unusual structure from forming (Fig.~\refSub{c}{fig-mw-atomic-nucl-snapshots}). Note that the structures shown in Fig.~\ref{fig-mw-atomic-nucl-snapshots} are early snapshots from the simulations: the entire structure freezes rapidly, but it is easier to see what is happening when the systems are not yet fully frozen.  We have classified brute-force simulations by their growth regime in a schematic plot shown in Fig.~\ref{fig-mw-atomic-alphabeta-schematic}; the three regimes identified correspond to the three snapshots shown in Fig.~\ref{fig-mw-atomic-nucl-snapshots}. The nucleation regime boundaries seen in Fig.~\ref{fig-mw-atomic-alphabeta-schematic} depend on both $\alpha$ and $\beta$, which illustrates that it is the interplay between two- and three-body interactions that determines the nucleation mechanism, rather than either one or the other on its own.

In all the `tight adsorption' cases where the two-body part of the potential is relatively more significant than the three-body part, the ice structure that grows is displaced from the underlying surface structure so that some atoms are in the `holes' in the middle of the surface chairs, which maximises their two-body interactions.\footnote{Of course in normal ice growth, such `stacking faults' also occur, but not for the same reason. The ice that grows in heterogeneous nucleation simulations with a fully ice-like structured surface is generally a mixture of cubic and hexagonal ice.} A schematic illustration of this bonding pattern is shown in  Fig.~\refSub{a}{fig-mw-atomic-nucl-snapshots} and Fig.~\refSub{b}{fig-mw-atomic-nucl-snapshots}; the two structures are different because in (a), there is a layer of water molecules fully penetrating the rigid surface layer and the first layer above the surface is considerably closer to the surface than in (b), which makes the energetic considerations different. In particular, the adsorption site that maximises the two-body interactions is the 6-co-ordinate site just above the centres of the sixfold rings, and the next best is the 4-co-ordinate site directly above the lower of the two types of surface particles; these two sites are occupied when $\beta$ is small. However, when $\beta$ increases, this makes the 6-co-ordinate site unfavourable because of the large number of non-tetrahedral bond angles with the surface particles, and the primary adsorption site shifts to the 4-co-ordinate site. The second adsorption site in this `mixed' regime is then above the centres of the sixfold rings of the surface. This site is favoured over the alternative position above the `high' surface atoms both because it affords more next-nearest neighbour interactions with the surface (three rather than just one) and because there are no unfavourable three-body interactions with the surface (in the alternative site, there is a non-tetrahedral angle involving the primary adsorption site and the `high' surface atom).

Given the behaviour we observe in relation to the parameters $\alpha$ and $\beta$, it may initially appear that our results are inconsistent with those of  Lupi and co-workers\cite{Lupi2014, Lupi2014b} and Singh and M\"{u}ller-Plathe,\cite{Singh2014} who simulated ice nucleation on a graphitic surface. In their simulations, there was no three-body interaction with the surface molecules at all, \textit{i.e.}~$\beta=0$. However, the two-body strength was considerably smaller than what we have considered here ($\alpha \approx 0.02$), and the surface penetration was avoided by the use of a considerably larger $\sigma$ for the surface-bulk interactions.

Another way in which we can avoid the first layer of water from penetrating into the surface layer is by placing an additional hard wall at the appropriate distance from the surface. Depending on where exactly we place this hard wall, we can compensate for the three-body interaction being too weak and crystallise ice with the normal structure.

We argued above that one of the reasons why a flat unstructured surface does not nucleate ice well is that the three-body terms are needed to provide a tetrahedral structure to the growing ice network. However, in simulations with a structured surface, we can grow ice even when $\beta=0$. The layer that forms at the surface is not ice-like in the sense that each water molecule has a larger number of nearest neighbours than there would be in ice; however, the reason that ice is nucleated at the surface in such simulations is that, although there is no three-body interaction with the surface itself, the positions of the molecules that can penetrate into the surface are controlled by the two-body interaction, which is both attractive and repulsive, depending on the interparticle distance. The surface therefore imposes the correct lattice parameter onto the penetrating water molecules, and the corrugation of the surface gives rise to a water layer that adopts an ice-like bilayer structure that mirrors the surface. Since these water molecules do have a three-body interaction with the remaining molecules, an ice structure grows on top of the first layer. This is the principal difference that allows a structured surface, even if it does not itself have three-body interactions, to facilitate ice nucleation, whilst a comparable flat, unstructured surface does not.

Finally, it is instructive to investigate the variation of the interfacial free energies of ice and liquid water in contact with the structured wall as a function of $\alpha$ and $\beta$ in the region where we see `direct' ice growth in brute-force simulations. We can only readily interpret our results when the ice nucleation does not involve layers that have a non-ice-like relation to the surface, as $\gamma_\text{wall-ice}$ cannot straightforwardly be calculated when such intervening layers are present, which limits the  range of $\alpha$ and $\beta$ in which the calculated values are meaningful. If $\beta$ is decreased significantly, the two-body term will dominate and the density next to the surface will then be very high, and if $\beta$ is too small, particles can penetrate into the surface `ice' structure, as shown in Fig.~\ref{fig-mw-atomic-nucl-snapshots}. However, this is generally only the case for simulations started from the liquid state, whereas a pre-formed block of ice in contact with the ice surface, as used in the interfacial free energy calculation simulations, will not typically interpenetrate the surface. Brute-force simulations can thus in principle result in the formation of ice-like structures rather different from those simulated in the interfacial free energy calculation simulations, making comparisons between the two approaches somewhat difficult. However, within the limited range of $\alpha$ and $\beta$ considered below, water in the brute-force simulations is `well-behaved' and does not penetrate the wall ice surface.

\begin{figure}[tbp]
\centering
\includegraphics{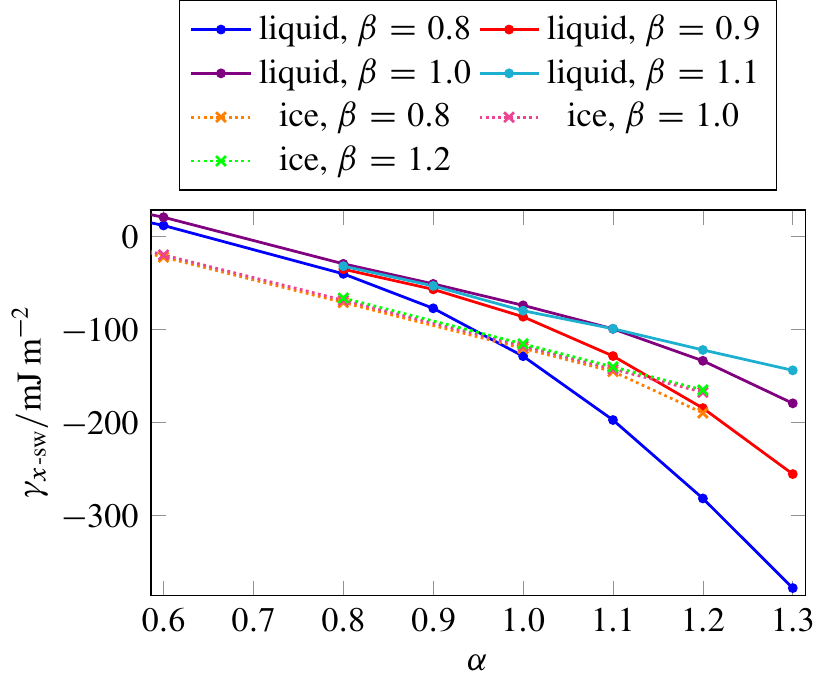}
\caption{The interfacial free energy of liquid water and ice in contact with a structured wall as a function of the two-body strength $\alpha$, for a selection of three-body strengths $\beta$. $T=\SI{273}{\kelvin}$, 360 surface particles.}\label{fig-mw-interfacial-free-energy-fn-two-three}
\end{figure}

The variation in interfacial free energies is shown in Fig.~\ref{fig-mw-interfacial-free-energy-fn-two-three}.\footnote{One potential difficulty when performing these simulations is that the liquid can sometimes freeze in simulations of the structured surface, particularly when the three-body strength $\beta$ is greater than unity. We do not include any structures that are detected by our order parameter as having crystallised (or partly crystallised) in the integration in $\lambda$, which means that several $U(\lambda)$ curves entail extrapolations of the liquid behaviour based on the smaller values of $\lambda$ at which freezing is not observed. This can lead to a decrease in accuracy, particularly for larger values of $\beta$. We have calculated several integrals in $\lambda$ in reverse (that is, starting from $\lambda=1$ and then gradually decreasing it), and while a small degree of hysteresis can be seen, the numerical answers obtained are the same within the simulation error.} There are several features to note about this plot. Firstly, when the two-body strength $\alpha$ is decreased, the interaction between the structured wall and both ice and liquid water naturally becomes less favourable; however, since it is the difference between the two interfacial free energies that determines the contact angle (Eqn~\eqref{eq:youngs-eqn}), the contact angle changes very little when $\alpha$ is reduced, as both the liquid and the ice curves are essentially linear with the same slope. By contrast, the reduction of the three-body strength $\beta$ by just \SI{20}{\percent} leads to a dramatic change in the contact angle: when we decrease the three-body interaction, the interfacial free energy becomes more favourable, as we are no longer penalising non-tetrahedral structures to the same degree, and this affects liquid water much more than it affects ice structures, which are in any case tetrahedral. When the two-body strength is decreased as well, this effect becomes less pronounced. For example, for the $\beta=0.8$ case, the surface becomes gradually less good at nucleating ice as $\alpha$ increases, consistent with what we observe in brute-force simulations. The converse arguments apply when two- and three-body strengths are increased. Finally, the ice structure can deform slightly if the three-body term is decreased too much relative to the two-body term; this is why the interfacial energy for the $\alpha=1.2$, $\beta=0.8$ point of ice is lower than the trend.

The variation of interfacial free energies with $\alpha$ and $\beta$ is entirely consistent with the brute-force simulations within the range of $\alpha$ and $\beta$ that can be studied with this method. Indeed, the decrease in the contact angle we see from interfacial free energy calculations  underlies the change in mechanism we see at lower values of $\beta$ in brute-force simulations.

\subsubsection{Varying the lattice parameter}

\begin{figure}[tbp]
\centering
\includegraphics{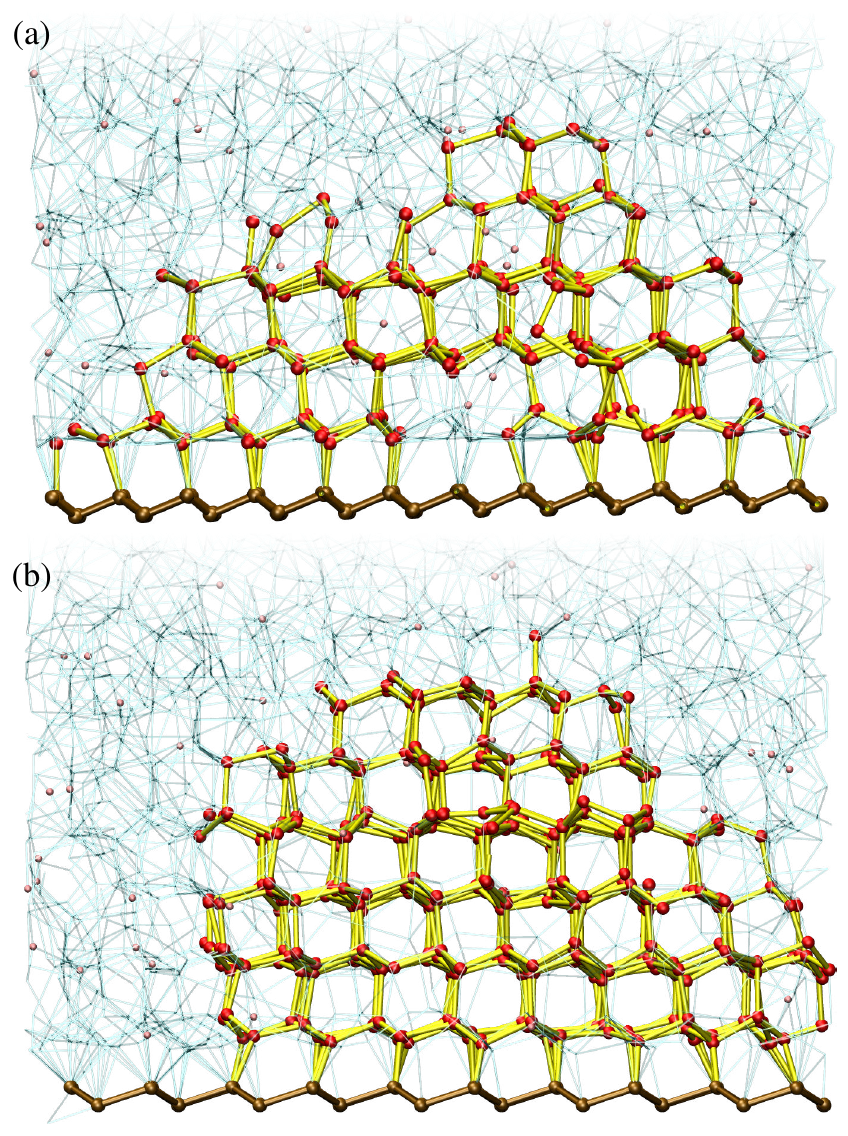}
\caption{Snapshots from simulations with a structured surface whose lattice parameters are (a) 0.86 and (b) 1.1 times that of ice at the simulation temperature. Only the region of the simulation box close to the surface is shown. $T=\SI{220}{\kelvin}$, $p=\SI{1}{\bar}$, $\alpha=1.0$, $\beta=1.0$. To make it easier to see the difference in the lattice parameter between the rigid surface and the growing ice nucleus, these snapshots are shown in an orthographic projection. (a) $N=2987$, of which 288 are fixed surface particles. (b) $N=2727$, of which 160 are fixed surface particles.}\label{fig-mw-atomic-nonComm-snapshots}
\end{figure}

Ice nucleation behaviour on a surface whose lattice parameter does not match that of ice can result in very interesting behaviour. It has, for example, been suggested that a `perfect' lattice match may not necessarily be the optimal one for the heterogeneous nucleation of ice\cite{Cox2012, Cox2013} or of Lennard-Jones particles on a crystalline surface.\cite{Mithen2014, *Mithen2014b} However, because the process is potentially fairly complex, the effects of changing the lattice parameter are not as unambiguous to rationalise as when we simply change the bond strengths, especially when the clusters are allowed to grow to reasonably large sizes.  If the lattice parameters are different from the equilibrium value, provided the structure is sufficiently large, this will result in defects at or near the surface in order to ensure that the ice near the surface can have the same density as bulk ice; this type of behaviour has been studied theoretically by Turnbull and Vonnegut.\cite{Turnbull1952} Furthermore, what happens during the nucleation process itself is not entirely obvious: as the growing nucleus becomes larger and the surface strain begins to build up, this might increasingly favour growth into the bulk rather than across the surface, thus changing the contact angle of the nucleus with the surface as it grows.

In particular, these considerations make it unfeasible to study lattice mismatch using the interfacial free energy approach we have used so far, but we briefly examine the nucleation behaviour as the surface lattice parameter is varied using brute-force simulations. We do not, however, analyse the nucleation rate as a function of the lattice parameter, and cannot therefore compare our results directly to those of Mithen and Sear.\cite{Mithen2014, *Mithen2014b}  We find that at \SI{220}{\kelvin}, there is a relatively rapid changeover from the surface being a good heterogeneous nucleant to it not nucleating ice growth significantly when the underlying surface lattice parameter decreases to less than about 0.86 times the equilibrium lattice parameter or increases to more than about 1.1 times the equilibrium lattice parameter. As an illustration, a snapshot from the simulation with a surface with a lattice parameter of 0.86 times that of the equilibrium lattice parameter is shown in Fig.~\refSub{a}{fig-mw-atomic-nonComm-snapshots}; we can clearly see that the nucleation is certainly no longer in the wetting regime, the first layer of ice that grows on the surface has a larger lattice parameter than the surface, and there are defects at the surface which allow ice clusters to grow to larger sizes. Similarly, for a surface whose lattice parameter is larger than bulk ice (Fig.~\refSub{b}{fig-mw-atomic-nonComm-snapshots}), the first layer of ice growing on the surface has a lattice parameter smaller than that of the surface. It seems that at these limiting values of the lattice mismatch, which are temperature-dependent, the strain introduced into the system is just below that which makes heterogeneous nucleation very slow and therefore difficult to observe in relatively short brute-force simulations.

\section{Conclusions}

We have investigated the behaviour of heterogeneous ice nucleation on a generic set of surfaces using a simple model of water. We find that the surface can influence the ice crystallisation pathway very considerably: certain surfaces do not facilitate ice growth, and ice nucleation can be homogeneous despite the presence of a surface, whilst other types of surface can act as excellent nucleants. More specifically, a flat attractive wall does not lead to an increased probability of nucleation relative to the homogeneous case, because the surface interacts more strongly with the denser liquid phase. By contrast, an ice-like basal surface that is lattice-matched is consistently able to nucleate ice even when the relative contributions of two- and three-body interactions are varied, albeit by different mechanisms in different regions of the interaction space. For this structured surface, it is only when the deviation from a lattice match reaches 10 to 15\;\% that there is a loss in the surface's nucleating ability; this is preceded by an increase in the contact angle and the presence of surface defects. For structured surfaces without a lattice mismatch, we have investigated the influence of two- and three-body effects on the interfacial free energies of ice and liquid water in contact with the surface, which in turn control the contact angle of the growing ice nucleus on the surface, and find that this contact angle can be increased significantly by reducing the tetrahedrality of the surface bonding or by increasing the interaction strength. This can result in a densification near the surface that can reduce the nucleation capacity of the surface, until eventually the mechanism by which heterogeneous nucleation occurs changes.

By considering a generic set of surfaces, we are able to formulate some general rules that a surface must fulfil in order to nucleate ice well (or, conversely, in order to be a poor nucleant). For example, our brute-force simulations have shown that it is not only the attraction to the surface that is important, but that orientational ordering, which in this case arises from the three-body interactions, is likewise crucial in order to achieve successful ice growth. However, we have also shown that, with a structured surface, if the two-body interaction is very strong, but the three-body interaction does not provide sufficient ordering of the first water layer, successful ice growth can nevertheless ensue, because an ice-like bilayer, albeit with too many neighbours, forms at the surface, and this then serves to nucleate the next layer in the structure. Interestingly, the formation of an adsorbed ice-like bilayer structure on the surface has been reported in simulations of water deposition on silver iodide;\cite{Shevkunov2005, *Shevkunov2005b, Taylor1993, *Hale1980, *Ward1982} we have seen analogous behaviour in preliminary simulations of \ce{AgI} in contact with bulk water. The formation of this bilayer may underlie silver iodide's excellent ice nucleating ability.

However, this behaviour can be contrasted to that seen by Cox and co-workers when studying heterogeneous nucleation on a structured surface\cite{Cox2012}  with surface atoms that are close-packed rather than in a corrugated ice-like arrangement. The flatness of their surface led to the formation of a flat layer of water molecules on top of it, which in turn resulted in a breakdown of the lattice match rule, because the first adsorbed layer had a topology different from that of ice when considering the lattice-matched surface.\cite{Cox2012} Perhaps unsurprisingly, the atomic arrangement of the underlying ice structure plays a very significant role in determining the heterogeneous nucleation mechanism.

While much remains to be learnt about heterogeneous ice nucleation, recent simulations have led to the development of a considerably clearer picture of the underlying physics of the process. Our work represents one approach to learning more about the general behaviour of water in contact with a surface; however, investigations of specific nucleants will be necessary to enable comparisons with experiments to be made and to begin to unravel the mysteries of heterogeneous nucleation in the atmosphere.

\begin{acknowledgments}
We thank the Engineering and Physical Sciences Research Council for financial support.
\end{acknowledgments}

\end{document}